\documentclass[a4paper]{aa}

\usepackage{graphicx}
\graphicspath{{./}{./Figures/}}
\usepackage{natbib}
\usepackage{amssymb}
\usepackage{amsmath}

\bibpunct{(}{)}{;}{a}{}{,} 

\def\mpc2{\ {\rm h}_{70} \ M_{\sun}.{\rm pc}^{-2}}

\newcommand{\be}{\begin{equation}}
\newcommand{\ee}{\end{equation}}

\newcommand{\adeu}{\mbox{ABCG$\,$209}}

\begin{document}

\title{Weak Lensing Mass Reconstruction of the Galaxy Cluster Abell 209
\thanks{This project has been partly supported by a Marie Curie
  Transfer of Knowledge Fellowship of the European Community's Sixth
  Framework Programme, under contract: MTKD-CT-002995 {\em COSMOCT}.}
}

\titlerunning{Mass Reconstruction of Abell 209}
\authorrunning{Paulin-Henrikson et al.}

\author{S. Paulin-Henriksson\inst{1}, V. Antonuccio-Delogu\inst{1}, C.P. Haines\inst{2}, M. Radovich\inst{2}, A. Mercurio\inst{2} and U. Becciani\inst{1} 
}

\institute{
  INAF / Osservatorio Astrofisico di Catania
  Via S. Sofia, 78, I-95123 Catania, Italy\\
\email{(sph,Vincenzo.Antonuccio,Ugo.Becciani)@oact.inaf.it}
\and
  INAF, Osservatorio Astronomico di Capodimonte, 
  Salita Moiariello 16, I-80131 Napoli, Italy
}

\date{Received ---; accepted ---}

\abstract
{
Weak lensing applied to deep optical images of clusters of galaxies 
provides a powerful tool to reconstruct the distribution of the
gravitating mass associated to these structures.
}
{
We use the shear signal extracted by an analysis of deep exposures of
a region centered around the galaxy cluster \adeu , at redshift \mbox{$z\sim 0.2$}, to derive both a map of the projected mass distribution and an
estimate of the total mass within a characteristic radius.
}
{
We use a series of deep archival \emph{R-band} images from CFHT-12k, covering an
area of $\sim 0.3\,$deg$^2$. We determine the shear of background
galaxy images using a new implementation of the modified
Kaiser-Squires-Broadhurst KSB+ pipeline for shear determination, which we 
has been tested 
against the
 ``Shear TEsting Program 1 and 2'' simulations. We use mass aperture
 statistics to produce maps of the 2 dimensional density distribution,
 and parametric fits using 
both Navarro-Frenk-White and singular-isothermal-sphere 
profiles to constrain the total mass.
}
{
The projected mass distribution shows a pronounced asymmetry, with an elongated
structure extending from the SE to the NW. This is in general agreement with
the optical distribution previously found by other authors. A similar
elongation was previously detected in the X-ray emission map, and in the
distribution of galaxy colours. The circular NFW mass profile fit gives a total mass of \mbox{$M_{200} = 7.7^{+4.3}_{-2.7}\,10^{14}\,{\rm M}_\odot$} inside the virial radius \mbox{$r_{200} = 1.8\pm 0.3\,$Mpc}.
}
{
The weak lensing profile reinforces the evidence for an elongated structure of \adeu , as previously suggested by studies of the
galaxy distribution and velocities. 
}

\keywords{Gravitational lensing: weak lensing -- Galaxies: clusters
    -- Clusters of Galaxies: individual (Abell~209)}

\maketitle


\section{Introduction}
\label{sec:introduction}
The study of the origin and structure of clusters of galaxies occupies
a central place in the current efforts to understand the origin and
evolution of the Universe. In particular, the determination of the
mass distribution of clusters can prove crucial for verifying the existence of dark matter in the Universe and eventually to determine its abundance and
dynamical evolution.\\
\noindent 
Cluster masses have traditionally been derived through the virial analysis
of the velocity dispersion of cluster galaxies, with the assumption of
dynamical equilibrium 
\citep[eg.][]{1998ApJ...506...45G,2001ApJ...548...79G}, and/or from the X-ray temperature of the hot
intracluster gas, assuming hydrostatic equilibrium \citep[][and references therein]{2002ARA&A..40..539R}. 
Since using these methods one has to
assume the dynamical and hydrostatic equilibrium, these mass
estimates are affected by the ignorance about the dynamical state of the cluster. Weak lensing analysis offers a unique opportunity to determine the cluster
mass distribution without such assumptions on its equilibrium, as the effect
is due to the gravitational deflection of the light that is dependent solely on the
distribution of matter. In particular, one application of weak lensing
analysis is the detection of the weak shear around galaxy clusters,
yielding an estimate of the total cluster mass and allowing a full
mass reconstruction of mainly low ($0.2\!\la\!z\!\la\!0.5$) redshift clusters \citep{2004ApJ...604..596C, 2005A&A...434..433B, 2006A&A...451..395C}. As a possible target for a weak lensing
analysis the galaxy cluster \adeu{} is particularly interesting, because the photometric and
evolutionary properties of its galaxy populations have been already
thoroughly studied 
\citep{2004A&A...424...79M,2003A&A...397..431M,2004A&A...425..783H}.\\
\noindent
In this paper we present the weak lensing mass reconstruction of the
galaxy cluster \adeu{} at $z=0.21$ \citep[][and references therein]{2003A&A...408...57M} 
using archival wide-field $R$-band imaging. \adeu{} 
is a rich \citep[richness class R=3,][]{1989ApJS...70....1A}, X-ray
luminous
\citep[L$_{\mathrm{X}}$\,(0.1-2.4\,keV)$ \sim 14\;h_{50}^{-2}\;10^{44}\,$erg.$s^{-1}$, ][]{1996MNRAS.281..799E}, moderately hot 
\citep[T$_{\mathrm{X}}\sim10$ keV,][]{1998MNRAS.301..328R} and massive cluster \citep{2002ApJS..139..313D,2003A&A...408...57M}. It is
characterized by the presence of substructures, which is shown by an elongation and asymmetry in the X-ray emission maps, with
two main clumps \citep{1998MNRAS.301..328R}. Moreover, the young dynamical state is indicated by the
presence of a radio halo 
\citep{1999NewA....4..141G, 2006astro.ph.10271V}, which has been suggested to
be the result of a recent cluster merger, through the acceleration of
relativistic particles by the merger shocks \citep{2002IAUS..199..133F}.\\

The plan of the paper is as follows. In section 2 we will shortly
review the data reduction procedures. In section 3 we will present the software
pipeline which we have developed to perform this analysis, 
based on the KSB+ algorithm \citep{1995ApJ...449..460K,
  1997ApJ...475...20L}. There exist many variants of this algorithm
\citep[see][for a presentation and comparison of some
  implementations]{2006MNRAS.368.1323H,2006astro.ph..8643M}, so we will describe in some
detail our particular implementation (the OACt pipeline). In section 4 we
will describe the preparation of the galaxy catalogue, and the technique
adopted to extract the shear information. Section 5 will be devoted to
present the mass estimated and a mass aperture reconstruction obtained
from the shear maps. In section 6 we will compare the mass maps with
the galaxy distributions and the X-ray maps. Section 7 will present 
the conclusions.


\section{Data description}
\label{sec:datadescription}

A detailed description of the data and reduction techniques has been
given elsewhere \citep{2004A&A...425..783H}, so here we will only
summarize the main steps. The data were obtained from the Canada-France-Hawaii telescope (CFHT)
science archive (PI J.-P. Kneib), and comprise a wide-field R-band image centered on the cluster. The observations were made on 14-16
November 1999, using the CFHT12K mosaic camera, an instrument made up of
12  $4096\times2048$ CCDs, set at the prime focus of the 3.6-m CFHT. The
CCDs have a pixel scale of  $0.206^{\prime\prime}$, resulting in a total
field of view of  $42\times28~{\rm arcmin}^{2}$, corresponding to
$8.6\times5.7~h^{-2}_{70}~{\rm Mpc}^{2}$ at the cluster redshift. The
total exposure time is 7200 s, made up of twelve 600s exposures, jittered to cover
the gaps between the CCDs.\\
\noindent
Standard IRAF tools were used to bias-subtract the images, using bias
exposures and the overscan regions of each CCD, before flat-fielding using
a superflat made up of all science images from the same observing run,
registering and Co-adding the images. The resultant images have a median
seeing of $0.73^{\prime\prime}$.\\
\noindent
The photometric calibration was performed in the the Johnson-Kron-Cousins
photometric system, using observations of $\sim$300 secondary standard
stars ($14<$R$<17$) in fields 6 and 7 of Galadi-Enriquez et al. (2000),
resulting in a zero-point uncertainty of 0.005 magnitude.\\
\noindent
For the weak lensing analysis we have masked: (i) saturated stars and
  hot pixels; (ii) a $1^{\prime}$ (ie. $300\,$pixels) border all
  around the field, where the point-spread function (hereafter PSF) is too
  complex to be properly modelled (eg. concave and/or varying too
  rapidly on small scales); (iii) CCD gaps (ie. area covered by
  several CCD on stacked images), where the PSF is also too complex.\\
\noindent 
The data set does also contain B-band imaging of the field with seeing and depth significantly lower than on the R-band image. On this B-band image, the bright star density is too low to interpolate the PSF, so it is not used to compute the shape parameters, but to help distinguish between cluster, foreground and background galaxies.
Using an algorithm which takes into account the R-band magnitude, B-R colour and the local
density, \citet{2004A&A...425..783H} attribute to each galaxy a
probability of belonging to the cluster. Given the lack of spectroscopic
information, this probability is the most accurate information we have
to discriminate among cluster and field galaxies.


\section{Weak gravitational shear estimate}
\label{sec:theory}
We extract the shear signal from observed polarisation of background
galaxies, corrected for the effects of the PSF, via the standard  KSB+ method \citep{1995ApJ...449..460K, 1997ApJ...475...20L}, improved as described in the following sections. 
We have blind tested our pipeline on the Shear-TEsting-Program (STEP)
simulated data \citep{2006MNRAS.368.1323H,2006astro.ph..8643M}, where
altogether 16 different weak lensing pipelines have been tested and compared. 
For reasonable PSFs, as the one sampled by the images presented in
this paper turned out to be, the shear we measure $\boldsymbol{\gamma}^{\rm
  meas.}$ is a linear function of the true (ie. simulated) one $\boldsymbol{\gamma}^{\rm true}$:
\be
\label{eq:biasdef}
\boldsymbol{\gamma}^{\rm meas.} = (m+1) \boldsymbol{\gamma}^{\rm true} + \boldsymbol{C}
\ee
where $m$ is the calibration bias and $\boldsymbol{C}$ is a systematic effect, mainly describing the anisotropy of PSF residuals. Both $m$ and $\boldsymbol{C}$ depend on the PSF, but contain a constant factor intrinsic to the pipeline, which can be subtracted. After subtraction, one has:
\be
\label{eq:biasmeasured}
\begin{array}{lll}
-0.05 & < m < & 0\\
-0.02 & < |\boldsymbol{C}| < & 0.02
\end{array}
\ee
For the current data set these systematics are well below statistical
errors. These statistical errors are mostly due to the intrinsic distribution of galaxy polarisations and to the background galaxy density.\\
\noindent
We present in sections \ref{sec:formalism} and \ref{sec:ksb} the basic
KSB+ formalism, and in section \ref{sec:oact} the details specific to
the OACt pipeline.


\subsection{Polarisation as a shear estimator}
\label{sec:formalism}
The observed polarisation of a galaxy offers an estimate of the local
gravitational shear, and can be defined using the weighted quadrupole
moments of the brightness distribution \citep{1995ApJ...449..460K}:
\begin{equation}
\label{eq:quadrupoles}
Q_{ij} = \frac{\int \, d^2\theta \, W(\theta) \,
  I(\theta) \, \theta_i
  \theta_j} {\int d^2\theta \, W(\theta) \,I(\theta) },
\end{equation}
where $I$ is the surface brightness of the object, $\theta$ is the angular
distance from the object center and $W$ is a window function, whoise
introduction is necessary
to reduce the shot noise to a reasonable level at large distances from
the centroid. Note that the indexes $i$ and $j$ are symmetric, so we have $Q_{12}=Q_{21}$. Using the weighted quadrupoles, we define the complex polarisation $\boldsymbol{e}=e_1+i\,e_2$ as:
\begin{equation}
\label{sec:defpolarisation}
\left(
\begin{array}{c}
e_1\\
e_2
\end{array}
\right)
= \frac{1}{Q_{11} + Q_{22}}
\left(
\begin{array}{c}
Q_{11} - Q_{22}\\
2Q_{12}
\end{array}
\right)
\end{equation}
For an elliptical object and a constant window function $W(\theta )=1$,
$\boldsymbol{e}$ is simply related to the axis ratio $\beta$. Defining
a position angle $\theta$ of the major axis, measured
counter-clockwise from the $x$ axis, one obtains:
\begin{equation}
\label{eq:elliparam}
\left(
\begin{array}{c}
e_1\\
e_2
\end{array}
\right)
= \frac{1-\beta^2}{1+\beta^2}
\left(
\begin{array}{c}
\cos 2\theta\\
\sin 2\theta
\end{array}
\right)
\end{equation}
In the weak lensing limit (defined by: $|\boldsymbol{\gamma}| \ll 1$, where $\boldsymbol{\gamma} = \gamma_1 + i\gamma_2$ is the complex shear field), and assuming that the real (ie. unobserved) polarisation distribution has a null average, $\boldsymbol{\gamma}$ is directly related to the average observed polarisation, $\boldsymbol{\gamma}\approx\boldsymbol{e}/2$.
 

\subsection{The KSB method}
\label{sec:ksb}
The current KSB+ method is the result of a series of successive
improvements \citep{1997ApJ...475...20L,1998ApJ...504..636H} of the
original method proposed by \citet{1995ApJ...449..460K}. It provides a
gravitational shear estimate by first-order subtraction of the PSF
smearing and shearing from the galaxy polarisation. 
The 2D vector KSB+ shear estimator of a single galaxy
\mbox{$\widehat{\boldsymbol{\gamma}}$} is given by:
\begin{equation}
\label{eq:shearestimator}
\widehat{\gamma}_\alpha = \left(P^\gamma \right)^{-1}_{\alpha
  \beta} \left[e_{\beta} - P^{\rm
  sm}_{\beta\mu}q_{\mu}\right].
\end{equation}
where we have adopted the standard convention on the summation rule of
indices, and $\boldsymbol{q}$ is the anisotropic component of the PSF, $\boldsymbol{P^{\rm sm}}$ is the smear polarisability tensor,
and $\boldsymbol{P^\gamma}$ is the pre-seeing shear polarisability
tensor, the latter being defined as:
\begin{equation}
\label{eq:pgammadef}
\boldsymbol{P^\gamma} = \boldsymbol{P^{\rm sh}}-\left(\boldsymbol{P^{\rm sm}}_{\rm PSF}\right)^{-1}\cdot\boldsymbol{P^{\rm sh}}_{\rm PSF}\cdot\boldsymbol{P^{\rm sm}}
\end{equation}
Here $\boldsymbol{P^{\rm sh}}$ is the shear polarisability tensor, and
the subscript ``PSF'' signals that the quantity is computed for the PSF. In equation \ref{eq:shearestimator}, $\boldsymbol{q}$ and
$\boldsymbol{P^\gamma}$ depend on the PSF and are estimated from the
images of the surrounding stars.\\
\noindent
The actual prescription to estimate $\boldsymbol{q}$,
$\boldsymbol{P^\gamma}$ and the PSF-subscripted tensors, the choice of the window function $W$ in equation \ref{eq:quadrupoles}, the algorithm of pixelised summations, and finally the approximations, vary from one implementation of the method to another. Our algorithm is described in the following section.


\subsection{The OACt pipeline}
\label{sec:oact}


\subsubsection{Smoothing radius, significance, Window function,centroid and Summation algorithm}
We determine the significance $\nu$ and the optimal smoothing
radius $r_g$ of each object. 
$\nu$ and $r_g$ are defined as usual in weak lensing: when convolving
the image with a Mexican-hat filter, the radius $r_g$ is the smoothing
radius for which  the object has the best signal-to-noise ratio, and
the  significance $\nu$ is this best signal-to-noise ratio.\\
\noindent
The window function $W$ of \mbox{equation \ref{eq:quadrupoles}} is
taken to be a circular Gaussian centered on the weighted centroid,
having a standard deviation equal to $r_g$. The weighted centroid, computed iteratively, is the point for which weighted dipoles are equal to zero:
\begin{equation}
\label{eq:dipoles}
\int \, d^2\theta \, W(\theta) \, I(\theta) \, \theta_i = 0
\end{equation}
Due to the finite size of pixels, all the integrals are replaced by
discrete sums with steps of $0.25\,$pixel in x and y directions,
truncated at a distance of $4r_g$. The flux at a given position is the
linear interpolation of the flux in the four nearest pixels, and the
background flux in a pixel is estimated by \textsc{se}xtractor (more details on
the actual parameters adopted are given in \mbox{section \ref{sec:alldet}}).


\subsubsection{Shape computation}
\label{sec:shapecomputation}
We compute the ellipticities, smear polarisability tensors and shear
polarisability tensors of background galaxies as described above (section \ref{sec:ksb}) . 
We do the same for stars, except that we do not use the smoothing
radii $r_g$ of stars themselves, but we perform the calculation for
every single $0.1\,$pixel-bin of $r_g$ in the range
$0.95<r_g<9.05$. This bin width of $0.1\,$pixel is chosen to be much
smaller than the accuracy of the measured $r_g$ (this accuracy is of
few tenths of pixel), and the range is given by extremum values of
$r_g$ in usual data. This gives a total of 81 $r_g$ bins.\\
\noindent
After computation of the shape parameters and shear estimator, we discard the objects for which: \mbox{(i) the} standard deviation on the centroid is larger than $0.2\,$pixel; \mbox{(ii) the} polarisation $|\boldsymbol{e}|$ is greater than 1; \mbox{(iii) the} shear estimator $|\boldsymbol{\gamma|}$ is greater than 2; \mbox{(iv) the} trace of the $\boldsymbol{P^\gamma}$ tensor is lower than 0.2. For the current data set, this corresponds to 5\% of the background galaxies. Even on noisier data sets this usually corresponds to less than 10\%. 


\subsubsection{Subtracting the smearing and shearing effects of the PSF}
The corrective factors of a given galaxy are all computed in the $r_g$
bin corresponding to the $r_g$ of the galaxy itself. In other words,
we interpolate PSF properties from stars in every single $r_g$ bin
independently for each galaxy. 
In all the $\boldsymbol{P^{\rm sm*}}$ and $\boldsymbol{P^{\rm sh*}}$
tensors of stars, the off-diagonal terms are completely dominated by
shot-noise, and typically they are more than one order of magnitude
smaller than the diagonal terms. For this reason, following similar
implementations \citep{2000ApJ...532..88H}, we systematically neglect these off-diagonal terms. 
Also the \mbox{non-diagonal} terms of the tensor $\boldsymbol{P^\gamma}$ are
extremely noisy, and we then approximate this tensor  
as a scalar equal to half its trace. 
These approximations imply that the vector $\boldsymbol{q}$ of
equation \ref{eq:shearestimator} (ie. the anisotropic component of the
PSF) is the interpolation of:
\begin{equation}
\label{eq:qoactdef}
q_\alpha = \frac{e_\alpha^*}{P^{\rm sm*}_{\alpha\alpha}}
\end{equation}
and :
\begin{equation}
\label{eq:pgammaoactdef}
P^\gamma = \frac{1}{2}\sum_\alpha\left(P^{\rm sh}_{\alpha\alpha} - \frac{P^{\rm sh*}_{\alpha\alpha}}{P^{\rm sm*}_{\alpha\alpha}}P^{\rm sm}_{\alpha\alpha}\right)
\end{equation}
Finally, we compute the shear estimator following this simplified version of equation \ref{eq:shearestimator}:
\begin{equation}
\label{eq:gammaoactdef}
\widehat{\gamma}_\alpha = \frac{1}{P^\gamma}\left[e_{\alpha} - \sum_i P^{\rm sm}_{\alpha i} q_i\right].
\end{equation}


\subsubsection{Mapping PSF properties}
\label{sec:mappingpsf}
From the star catalogues, we interpolate the 4 ratios:
$(e_\alpha^*\,/\,P^{\rm sm*}_{\alpha\alpha})$ and $(P^{\rm
  sh*}_{\alpha \alpha}\,/\,P^{\rm sm*}_{\alpha \alpha})$ (with $\alpha=1$ or 2) over the entire 
region (where the asterisks refer to parameters measured on
stars). The former two terms give $\boldsymbol{q}$ (equation
\ref{eq:qoactdef}) and the latter two give $\boldsymbol{P^\gamma}$
through equation \ref{eq:pgammaoactdef}. For each term, we fit a
2-dimensional, 2-degree polynomial independently on each CCD (and also independently in each $r_g$ bin).\\
\noindent
As we will see in the following, in order to keep a
relatively high star density, we use a
rather permissive star selection criterion. As a consequence, our star
catalogue is contaminated by small galaxies. In order to reject these small galaxies when fitting PSF properties, the fits are iterative: after the first fit is performed, we
reject all objects with at least one residual at more than $3\sigma$,
and we continue to iteratively perform new fits until
convergence. Typically, this procedure converges after 2 or 3 iterations.


\section{Detection of the weak lensing signal}
\label{sec:polarisationmeasurements}
To extract the weak shear information from the reduced and calibrated CHFT12k images presented in section \ref{sec:datadescription}, we first build the star-field and the background-galaxy-field catalogues. Background galaxies contain the weak lensing signal, smeared and sheared by the PSF, while stars are measures of the local PSF. 
We first detect all the objects within the field and then select those
relevant for weak lensing. Stars are selected according to their sizes
and their magnitudes, while the background galaxies are selected by
cross-checking our relevant object catalogue with the galaxy catalogue
of \citet{2004A&A...425..783H}. The latter contains all galaxies
within the field and assigns to each a probability of belonging to the field rather than to the cluster itself. All these steps are described in section \ref{sec:alldet}. 
We then compute shape parameters of stars and background galaxies (section \ref{sec:shapecomputation}), using the stars to map the PSF (as described in section \ref{sec:mappingpsf}), and finally compute a shear estimator for each background galaxy.


\subsection{Star and background galaxy catalogues}
\label{sec:alldet}
\begin{figure}
  \centering
  \includegraphics[scale=0.5]{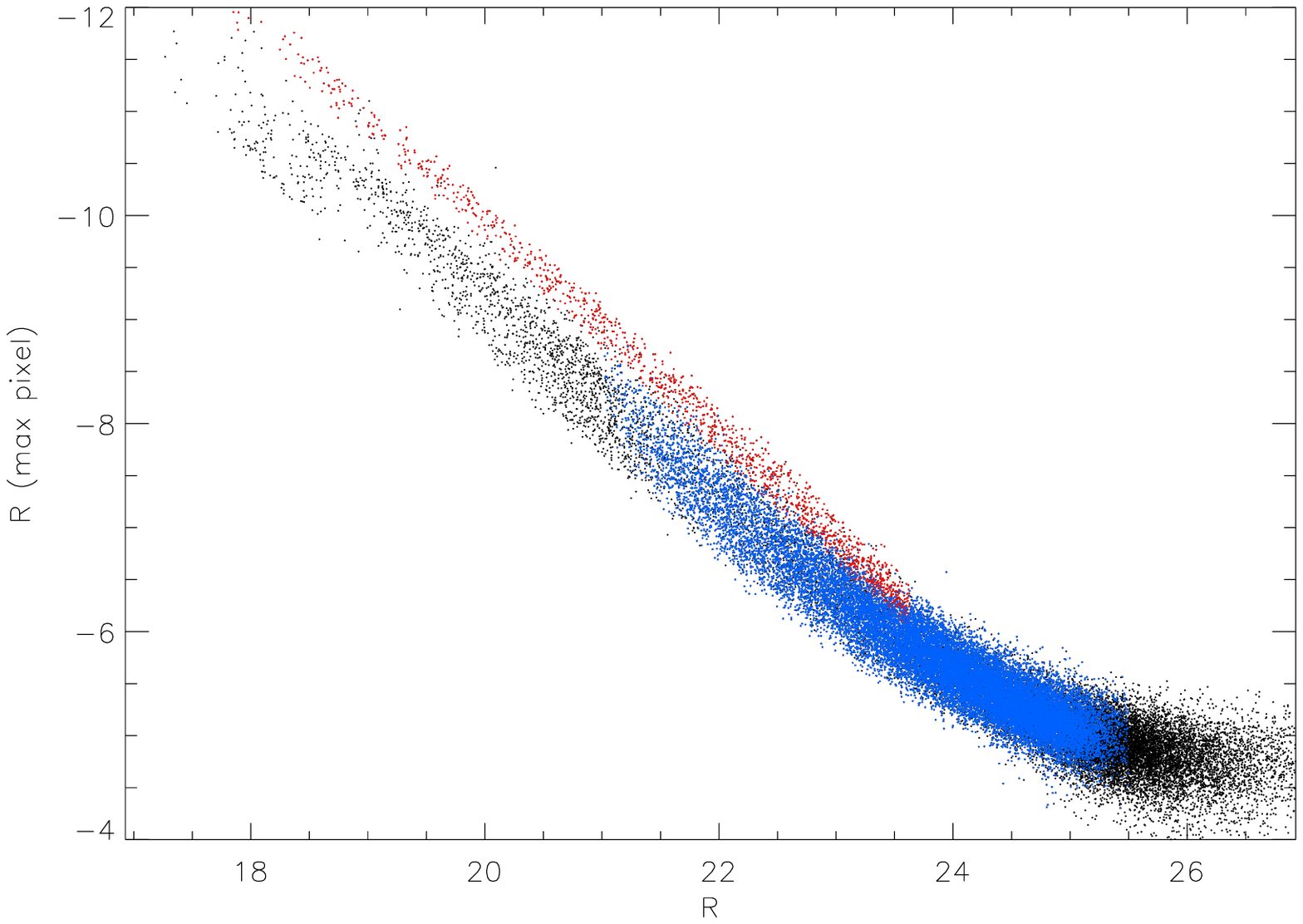}
  \includegraphics[scale=0.5]{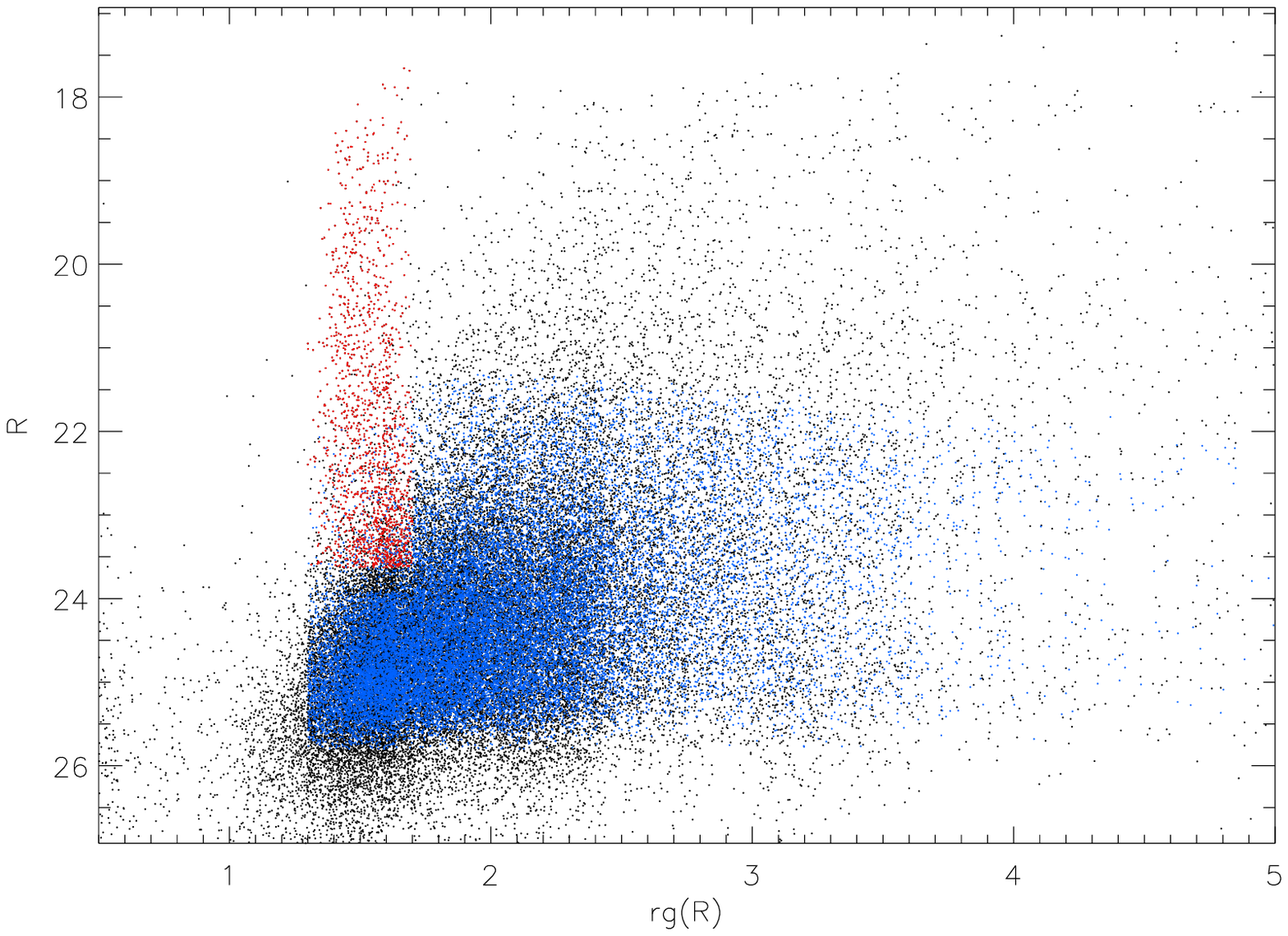}
  \caption{
Top (bottom respectively) panel: magnitude - central magnitude (optimal smoothing radius $r_g$ - magnitude resp.) diagram for all objects detected in the image. Objects selected as possible stars are in red, those selected as possible background galaxies are in blue. Other objects are in black.
}
  \label{fig:stargalsepr}
\end{figure}

We detect all the objects on the image using \textsc{se}xtractor, with very low thresholds. We also get a large proportion ($\sim 50\%$) of spurious detections but these are rejected later. We demand detected objects to have at least 5 pixels above $1.5\,\sigma$, where $\sigma$ is the standard deviation of the local background (ie. \textsc{detect\_thresh} and \textsc{analysis\_thresh} set to 1.5, \textsc{detect\_minarea} set to 5), where the local background is estimated with keywords \textsc{back\_size} set to 70 and \textsc{back\_filterize} set to 5. 
\textsc{se}xtractor is also used 
to measure the flux and the half-light-radius $r_h$ of each object. 
We then compute $r_g$ and $\nu$, as described in section \ref{sec:oact}. 
We then successively remove from the \mbox{catalogues :}
\mbox{(i) the} objects with $\nu <3$; \mbox{(ii) the} objects with at
least one neighbor nearer than $3(r_g+r_g(\textrm{neighbor}))$;
\mbox{(iii) the} objects with at least one pixel belonging to the
masked area within an aperture of $3\,r_g$ (the masked area is 
described in section \ref{sec:datadescription}); \mbox{(iv) the}
objects with $r_g$ lower than the local smoothing radius of
stars. Finally, we compute the shapes of the remaining objects and
apply shape cuts, as described in section
\ref{sec:shapecomputation}. After having performed these steps, we
have a catalogue of $\sim 30000$ detections, containing all the
weak-lensing-relevant background galaxies but also stars, cluster
galaxies and a large proportion of spurious events due to the low
thresholds used with \textsc{se}xtractor.\\
\noindent 
We build a star catalogue with a loose selection based on 5 parameters: 
the significance $\nu$; the magnitude $R$; the surface brightness in
the central pixel $R_{\rm max}$; the half-light-radius $r_h$, and the
smoothing radius $r_g$. We demand: \mbox{$\nu>10$}; \mbox{$R\le
  24.0$}; \mbox{$R+2.6\le R_{\rm max}\le R+3.3$};
\mbox{$1.9<\frac{r_h(R)}{1\,{\rm pixel}}<2.5$};
\mbox{$1.3<\frac{r_g(R)}{1\,{\rm pixel}}<1.7$}. This returns
$1588\,$objects (\mbox{$\sim 1.5\,$objects.arcmin$^{-2}$}) uniformly
distributed in the field. We define this catalogue {\em loose} because
$\nu$, $r_g$ and $r_h$ of stars show variations of $\sim 30\%$
through the field with the PSF (while we apply these constant
cuts). Thus, these cuts result in a star catalogue which contains
$20\,-\,45\%$ of non-stellar detections, most of them being small
galaxies. These fake stars are iteratively rejected during the PSF
property fits, as described in section \ref{sec:mappingpsf}. At the
end of iterations, we are left with \mbox{$\sim 0.9\,$stars.arcmin$^{-2}$}, as shown in figure \ref{fig:starellmapr}.\\
\noindent
We select background galaxies from the remaining objects in two
steps. First, we reject fake detections and cluster galaxies by
cross-checking the remaining objects with the catalogue of
\citet{2004A&A...425..783H}, which includes all the galaxies within the
field, together with their accurate photometry and the probability for
each galaxy of belonging to the field rather than to the cluster itself
(see section \ref{sec:datadescription}). We include in our final
catalogue those galaxies identified in the catalogue of
\citet{2004A&A...425..783H} marked as having a probability larger than 80\% to belong to the field. Second, we reject most of the foreground galaxies with the cut $R>21$ (since the R-band magnitude is the only information we have at this stage). In order to reject too faint background galaxies for which the shape parameters are not reliable, we cut the $\sim 3\%$ faintest galaxies for which $R>25.5$.\\
\noindent
These cuts are optimised to give a background galaxy catalogue with low foreground contamination, while keeping almost all the relevant
background galaxies. They are illustrated in figure
\ref{fig:stargalsepr}, where the star sequence is clearly visible in
red, while background galaxies are in blue. The final background
galaxy catalogue contains $16708\,$galaxies (\mbox{$16.7\,$galaxies.arcmin$^{-2}$}). 
When combining individual galaxy shears to produce shear maps, mass
reconstructions or density profile fits (see sections
\ref{sec:shearmaps} and \ref{sec:massdistrib}), we weight galaxies
according to their significance, as proposed by
\citet{2006A&A...451..395C}: we do not consider objects with $\nu<5$,
while for $\nu>5$ the weight is set equal to ${\rm
  min}(\nu\,;\,40)$. In other words, for a given galaxy $i$, the
weight $w_i$ is defined as:
\begin{equation}
\label{eq:weight}
\begin{array}{lll}
w_i = 0 & {\rm ; if } & \nu_i<5\\
w_i = \nu_i & {\rm ; if } & 5\le\nu_i\le 40\\
w_i = 40 & {\rm ; if } & \nu_i>40
\end{array}
\end{equation}
The weighted galaxy number is $\Sigma_{i}w_{i}/{\rm max}(w_{i}) =
\Sigma_{i} w_{i}/40 = 8816$ (\mbox{$8.8\,$weighted galaxies.arcmin$^{-2}$}).


\subsection{Testing the PSF subtraction}
\label{sec:testingpsfsub}

\begin{figure}
  \centering
  \includegraphics[scale=0.5]{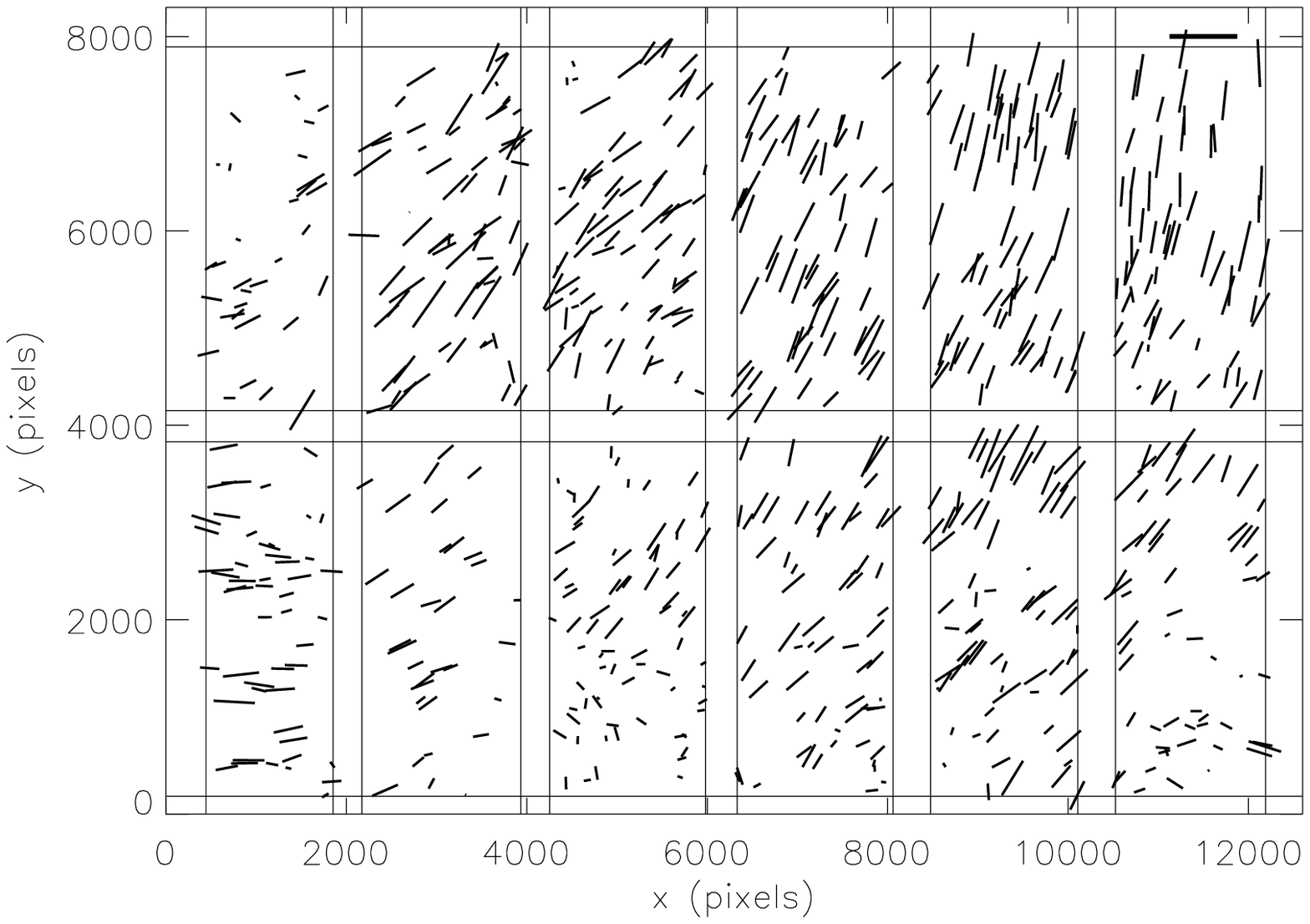}
  \includegraphics[scale=0.5]{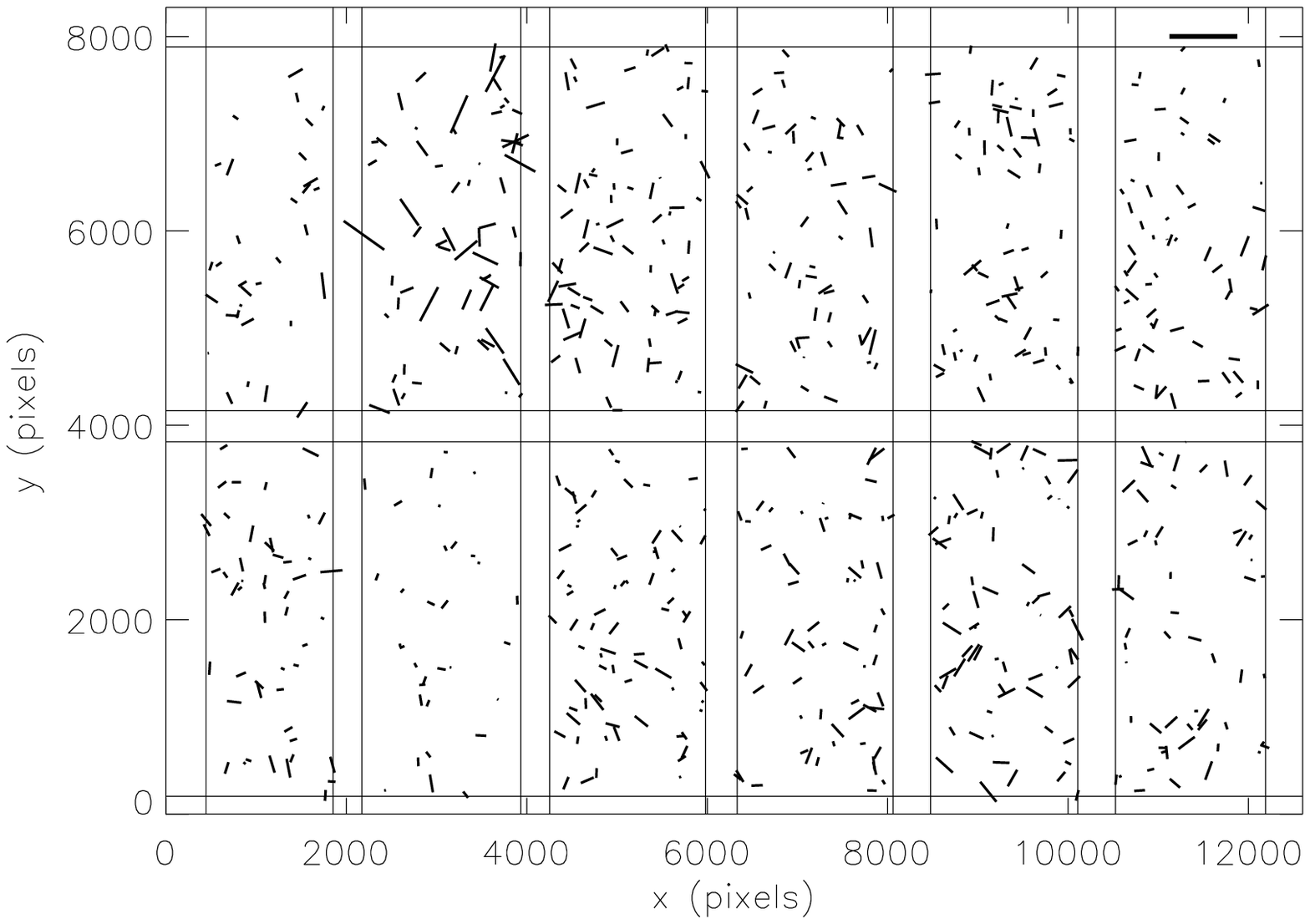}
  \caption{Star polarisations in the field. Top panel: before any
  correction (ie. measured polarisations). Bottom panel: after the
  anisotropic correction. In both panels, the vectors are oriented
  along the major axis of the ellipsoid, their length being
  proportional to the polarisation:
  \mbox{$|\boldsymbol{e}|=\sqrt{(Q_{11}-Q_{22})^2+4\,Q_{12}^2}\,/\,(Q_{11}+Q_{22})$}.
  The scale shown in the upper right corners is
  \mbox{$|\boldsymbol{e}|=0.03$}. The straight lines show the (masked) regions covered by several unstacked images.}
  \label{fig:starellmapr}
\end{figure}

\begin{figure*}
\begin{center}
  \includegraphics[scale=0.7]{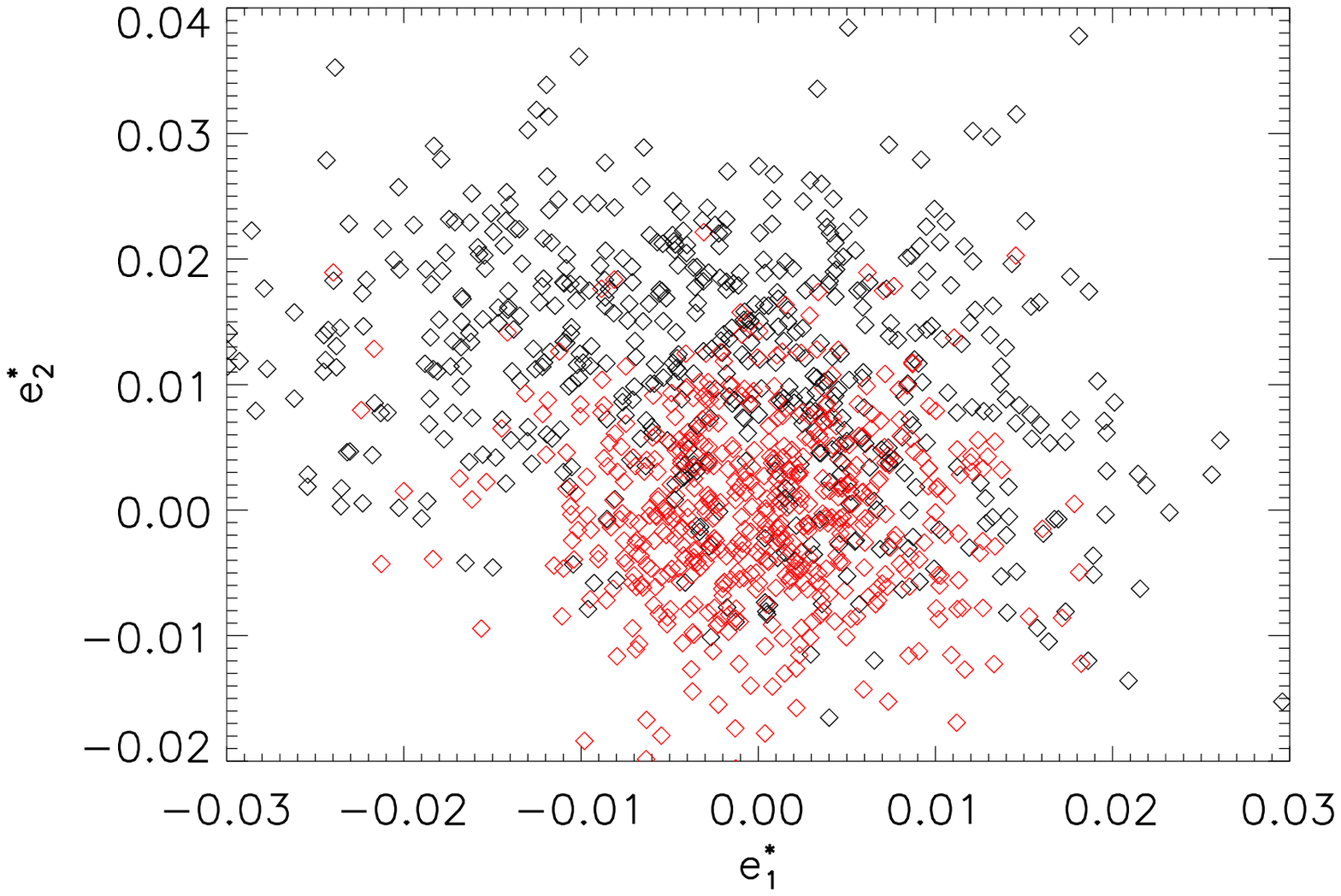}
  \includegraphics[scale=0.7]{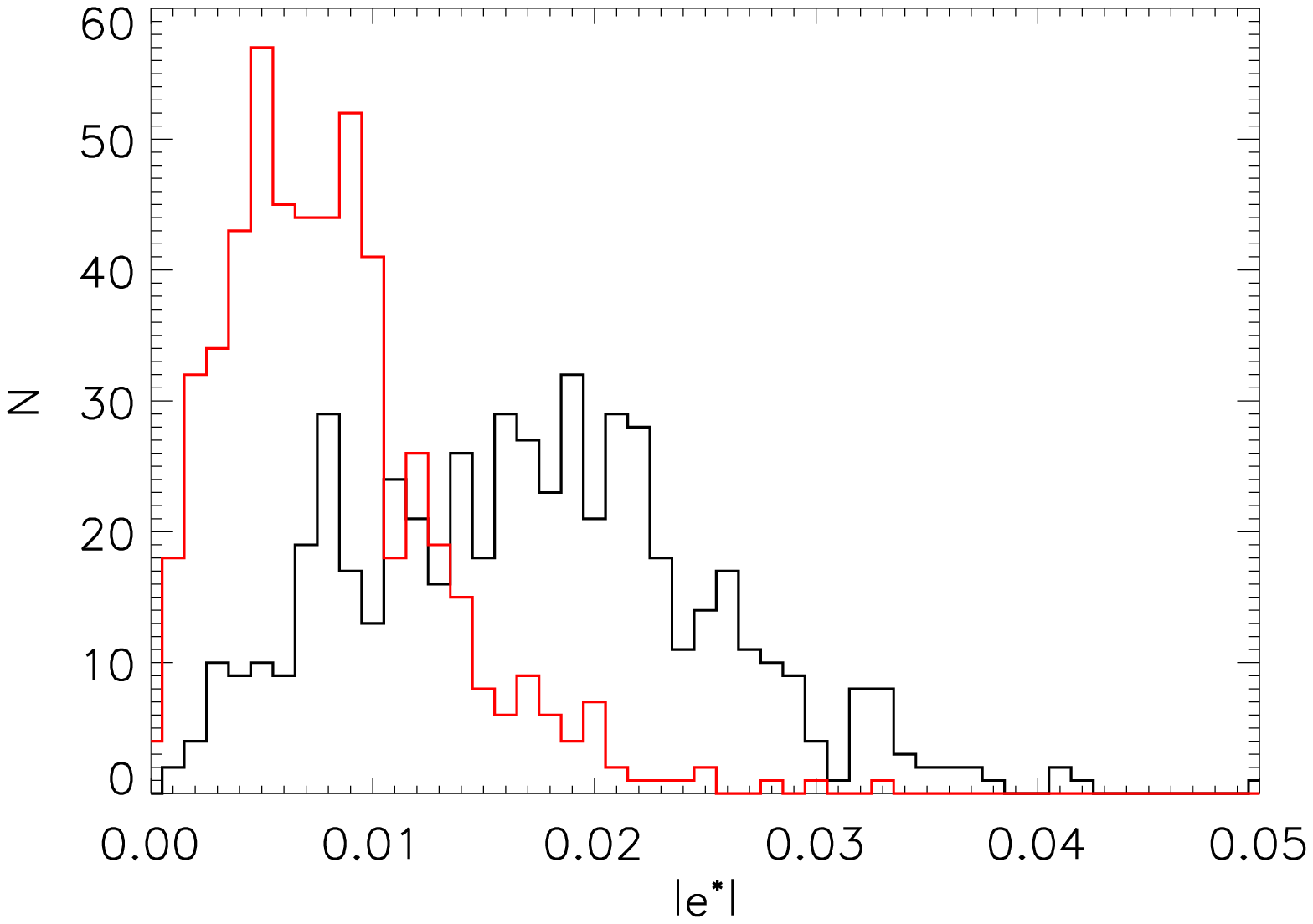}
  \caption{Distribution of stellar polarisations. Upper panel: $e_1$ against $e_2$ of stars before (black) and after (red) the anisotropic correction. Bottom panel: $|\boldsymbol{e}|$ distribution (same colours).
}
  \label{fig:epsstarsbeforeafter1}
\end{center}
\end{figure*}

Figure \ref{fig:starellmapr} shows the star polarisations over the
field before and after the anisotropic correction (ie. the subtraction
of $\sum_i P^{\rm sm}_{\alpha i} q_i$ in equation
\ref{eq:gammaoactdef}). One can notice the lack of large scale
correlations due to the PSF anisotropies, after the subtraction. The
correction also reduces the average amplitude and anisotropy of the PSF, as
it is clear from Figure \ref{fig:epsstarsbeforeafter1}. The corrected
distribution of ellipticities is more isotropic, and has also a smaller
scatter dispersion.


\subsection{Combining the individual shears, building shear maps}
\label{sec:shearmaps}
\begin{figure*}
\begin{center}
  \includegraphics[scale=0.9]{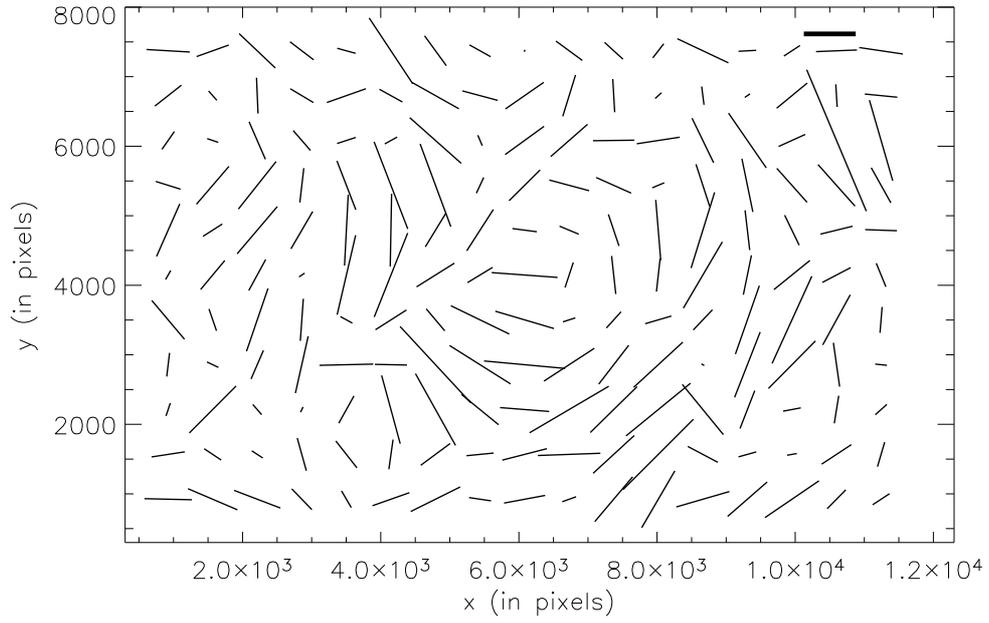}
  \caption{Shear map of the whole field obtained with an overlap of
  50\% (ie. 50\% of the field area is accounted into more than one cell). The scale is given by the thick line in the upper right corners : $\gamma = 0.05$.}
  \label{fig:shearmaps}
\end{center}
\end{figure*}

To get a shear map, we divide the field into $17\times 11$ square cells
with an overlap of 50\% (ie. 50\% of the galaxies in one cell belong only to this cell, while
the remaining 50\% belong also to at least one of the 8 neighbouring
cells). In each cell we average
$\boldsymbol{\gamma}$ according to the weighting scheme described in
section \ref{sec:alldet}. The resulting shear map is shown in Figure
\ref{fig:shearmaps}. One can see a characteristic pattern of increased
tangential shear, which coincides with the central region of the
cluster, as defined by the optical distribution of galaxies. This
visual impression is confirmed by the mass aperture map, as we will
see in the following sections.

\clearpage


\clearpage


\section{Mass distribution}
\label{sec:massdistrib}
There are quite a few different methods to deduce the mass
distribution from the individual shears of background galaxies \citep[see eg.][for a
  review]{2001PhR...340..291B}. A class of these methods makes use
of the 2D {\em smoothed} shear maps to get the projected surface density
distribution: examples of this class are the {\em mass
  reconstruction} method \citep{1995A&A...294..411S, 1995A&A...297..287S,
  2001A&A...374..740S}, or the {\em mass aperture} technique
\citep{1996MNRAS.283..837S}. These methods suffer from the {\em sheet
  mass degeneracy}: they provide a reliable reconstruction of the surface density, but this
is known except for a constant value. The alternative is to assume
{\em a priori} a given mass profile, so that the degeneracy is removed,
then attempt a parametric fitting. These latter methods are useful to get an
estimate of the total mass of the cluster. Both methods suffer from
possible systematic effects, for instance the contamination from
background sources.\\
\noindent
In section~\ref{sec:modelfitting}, we fit two different parametric
spherical density profiles: a Singular Isothermal Sphere
(SIS) and a Navarro-Frenk-White (NFW) \citep{1996ApJ...462..563N}. In section \ref{sec:massap}, we map the mass distribution using mass aperture statistics.

In order to derive actual values of the mass from the shear parametric, we have to know the critical surface density:
\be
\label{eq:sigmacritdef}
\Sigma_{\rm crit} = \frac{c^2}{4\pi G} \frac{\left\langle D_{\rm{ls}} / D_s \right\rangle^{- 1}}{D_{\rm{l}}}
\ee
where  $D_{\rm ls}$, $D_{\rm l}$ and $D_{\rm s}$ are the lens-source,
observer-lens and observer-source distances respectively. We do not
have however any redshifts of the background galaxies, so we approximate
their redshift distribution by a single median value (ie. we adopt a
source plane approximation). As described in section \ref{sec:alldet}, we restrict the background galaxies to the range \mbox{$21 < R < 25.5$}: in this magnitude
range, we can assume the median redshift to be $z \sim 1$ \citep{2004A&A...422..407G}. Then the critical density at the redshift of \adeu{} ($z=0.21 $) becomes:
\be
\label{eq:sigmacritval}
\begin{array}{lll}
\Sigma_{\rm crit} & \simeq & 1.62\times 10^{14}\,{\rm M}_\odot .{\rm arcmin}^{-2}\\
 & \simeq & 3.91\times 10^{15}\,{\rm M}_\odot.{\rm Mpc}^{-2}
\end{array}
\ee
for the basic $\Lambda$CDM cosmological model derived from the 3-year WMAP data \citep{2006astro.ph..3449S}:
\mbox{$\Omega_{M} = 0.27$}, \mbox{$\Omega_{\Lambda} = 0.73$},
\mbox{${\rm H}_0 = 70\,$km.sec$^{-1}$.Mpc$^{-1}$} and \mbox{$w=-1$}.\\


\subsection{Parametric model fitting}
\label{sec:modelfitting}

For both profiles we have performed a $\chi^{2}$ minimization of the
tangential shear:
\be
\label{eq:chisqdef}
\chi^{2} = \sum_{i=1}^{N} w_{i}\left(\gamma^{T}_{i} - \gamma^{T}_{\rm model}(r_{i})\right)^{2}
\ee
where $\gamma^T$ is the tangential shear (ie. the
shear projected along the direction orthogonal to the line connecting
the galaxy position to the cluster center), $w_{i}$ is the weight (as
defined in equation~\ref{eq:weight}) and $r_i$ is the distance to the
cluster center of mass. The latter is defined as the point for which the signal-to-noise of the $\gamma^T$ profile attains its maximum value:
\be
\label{eq:masscenterdef}
\frac{S}{N}(\alpha ,\delta )= \sum_{i=1}^N w_i\,\gamma^T_i(\alpha ,\delta )
\ee
By maximising eq.~\ref{eq:masscenterdef}, we find the center at \mbox{$\alpha=1$h$31$m$51.6s$} ; $\delta=-13^\circ 36^\prime$, 
offset by about $36^{\prime\prime}$ from the cD galaxy, as shown on figures \ref{fig:galdist}.\\
\noindent
For the NFW fit we have either assumed that
each galaxy provides an estimate of the underlying shear field, (so
the summation in equation~\ref{eq:chisqdef} extends to the full sample) or we
have binned the data and minimised over the averages in the bins. We
have verified that both procedures give compatible results.\\
\noindent
We exclude from the fits galaxies lying within an inner central 
region of the cluster, where the weak lensing approximation ($\kappa <
1$) is not everywhere valid, and outside an outer region, where the
statistics are very weak.\\
\noindent 
The actual fitted region lies within the following bounds: \mbox{$1\arcmin < \theta < 10\arcmin$} (\mbox{$0.2\,{\rm Mpc} < R < 2\,{\rm Mpc}$}) around the cluster center of mass. 


\subsubsection{Fitting a NFW profile}

The NFW profile has often been used as a good fit of numerically
simulated halos \citep{1995MNRAS.275...56N,1997ApJ...490..493N}. 
Although this fit was originally made only for simulated halos in
standard CDM models, it turned out to be a good fit for 
halos which formed in $\Lambda$CDM models. 
The mass density of the NFW profile is described by:
\begin{eqnarray}
  \rho(r) & = & \frac{\delta_{\rm c} \rho_{\rm c}}{(r/r_{\rm
      s})(1+r/r_{\rm s})^2} \\
  \mbox{where:} \,\,\, \delta_{\rm c} & = & \frac{200}{3} \frac{c^3}{\ln(1+c)-c/(1+c)} \\
  \mbox{and:} \,\,\,\rho_{\rm c} & = &
  \frac{3 H^2(z)}{8 \pi G}\,.
\end{eqnarray}
In the equations above $r$ is the distance to the cluster center, $r_{\rm s}$ is the scale radius, $H(z)$ the Hubble parameter at the
redshift of the cluster and \mbox{$c = r_{200}/r_{\rm s}$} is the
concentration parameter, where $r_{200}$ is the virial radius. 
The NFW density profile is
shallower than the SIS profile near the center but steeper in the outer parts. The total mass inside the radius $R$, shown on figure \ref{fig:fittedmass}, is:
\be
\label{eq:mrnfw}
M(<R) = 4\pi\,\delta_c\,\rho_c\,r_s^3\,\left(\ln (1+R/r_s) -1 + \frac{1}{1+R/r_s} \right)
\ee
Exact expressions for the tangential shear due to this mass distribution are given by
\citet{2000ApJ...534...34W}. It has two
highly correlated degrees of freedom: $r_s$ and $c$. From a \mbox{non-linear} 
\mbox{least-squares} fit (Levenberg-Marquardt) we obtain:
\be
\label{eq:resultnfwfit}
\begin{array}{lll}
c & = & 3.4^{+3.1}_{-1.6}\\
r_{\rm s} & = & 0.50^{+0.60}_{-0.25}\,{\rm Mpc}\\
\end{array}
\ee
This corresponds to virial radius $r_{200}$ and mass $M_{200}$:
\be
\begin{array}{lll}
r_{200} & = & 1.81^{+0.30}_{-0.26}\,{\rm Mpc}\\
M_{200} & = & 7.7^{+4.3}_{-2.7}\,10^{14}\,{\rm M}_\odot
\end{array}
\ee


\subsubsection{Fitting a SIS profile}
\begin{figure*}
  \centering
  \includegraphics[scale=0.7]{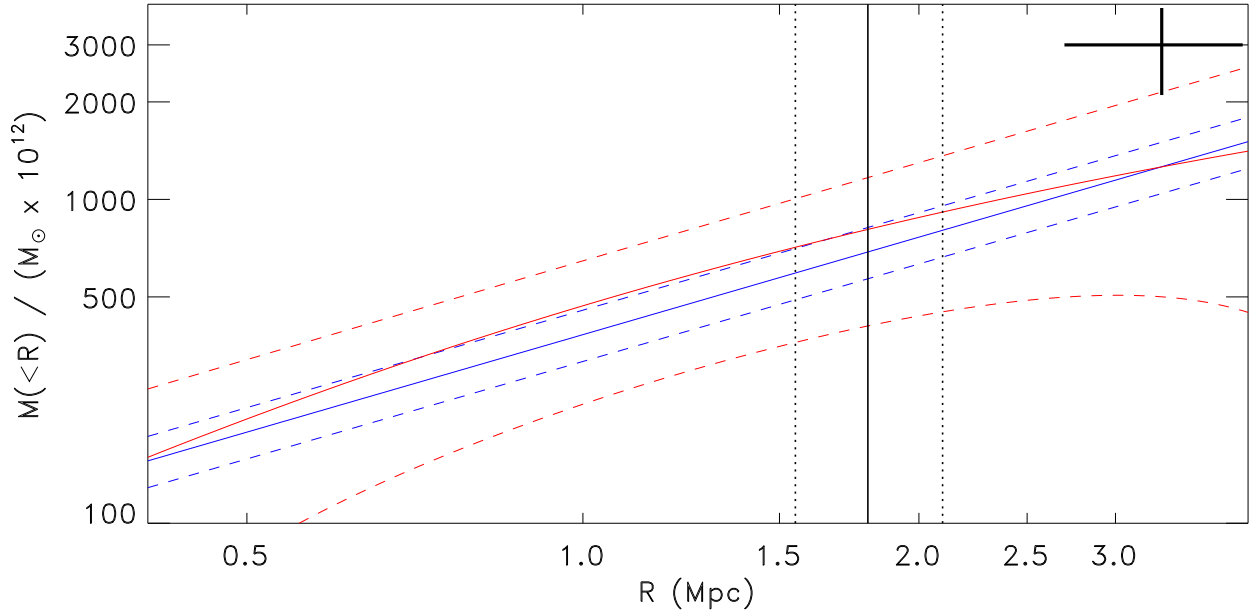}
  \caption{
Total mass inside the radius $R$ for the two models SIS (in blue) and NFW (in red). 
Dashed lines are the 66\% confidence level regions. 
The black vertical lines show the virial radius $R_{200}=1.81^{+0.30}_{-0.26}\,{\rm Mpc}$ given by the NFW fit: straight line for the best fit and dashed lines for the 66\% confidence level region. The black cross (in the upper right corner) shows the measure of $M(<R_{vir})$ and $R_{vir}$ by \cite{2003A&A...397..431M} from internal dynamics, discussed in section \ref{sec:compargaldistrib}.
}
  \label{fig:fittedmass}
\end{figure*}

For a SIS model, the density and shear profiles are respectively given by:
\begin{eqnarray}
\rho(r) & = & \frac{\sigma^2}{2\pi G r^2} \\
\kappa(\theta) & = & \gamma^T(\theta) = \frac{4\pi\sigma_v^2}{c^2}\frac{D_{\rm ls}}{D_{\rm S}}\frac{1}{2 \theta}
\end{eqnarray}
where $\sigma_v$ is the velocity dispersion, the only free parameter
in this model. The total mass inside the radius $R$ (shown on figure \ref{fig:fittedmass}) is:
\be
\label{eq:mrsis}
M(<R) = \frac{2}{G}\,\sigma_v^2\,R
\ee
A linear least-squares fit provides:
\be
\label{eq:resultsisfit}
\sigma_v = 924\pm 84\,{\rm km.s}^{-1}
\ee
The previous analysis from \citet{2002ApJS..139..313D} gives a
significantly lower value (\mbox{$\sigma_v = 680_{-130}
  ^{+120}\,$km.s$^{-1}$}), but the authors adopt a less restrictive
criterion to eliminate cluster galaxies before the weak lensing
analysis, so their weak lensing signal, and consequently $\sigma_v$,
could be underestimated. 
Although the total mass of a SIS model diverges, we can however show
the integrated mass inside a given radius $R$ (Figure
\ref{fig:fittedmass}). At the virial radius $R=r_{200}\sim
1.8\,$Mpc, as provided by the NFW profile fit (previous section), the
mass inside the radius is almost the same both for SIS and NFW profiles: \mbox{$M(<R)= M_{200}\sim 7.7\times 10^{14}M_\odot$}.\\
\noindent
In Figure \ref{fig:fittedmass} we show the total mass inside the
radius $R$, according to the two different fits we find. Note that the $\pm 1\sigma$ region of NFW is much larger than in the SIS case. This is a direct consequence of equations \ref{eq:mrnfw} and \ref{eq:mrsis}: in the SIS model $M(<R)$ is proportional to $\sigma_v^2$ and so the confidence region has a width of $\sim\pm 20\%$, while in the NFW model $M(<R)$ depends on $\delta_cr_s^3f(r_s,R)$. This makes the $\pm 1\sigma$ region to have a minimum $\sim\pm 50\%$ width for $R\sim 1.3\,$Mpc.


\subsection{Mass aperture statistics}
\label{sec:massap}

The mass within an aperture radius $r$ can be obtained directly from the shear 
using the {\em aperture densitometry} statistics $\zeta$, defined as
in \citet{1994ApJ...437...56F}:
\begin{eqnarray}
\label{eq:massap1}
\zeta (R_1, R_2) & = & \bar{\kappa} (R_1) - \bar{\kappa} (R_1, R_2)\\ 
 & = & \frac{2}{1 - R^2_{^{} 1^{}} / R^2_{^{} 2^{}}} \int^{R_2}_{R_1}\left\langle \gamma_T \right\rangle_{r}\,{\rm d}\, \ln r
\end{eqnarray}
The mass inside the annulus defined by $(R_1, R_2)$ is connected to the
above quantity by: $M_{\rm ap} = \pi r^2 \zeta  \Sigma_{\rm
  crit}$. This is in fact a lower limit on the true mass, but only dependent on the tangential shear. Thus, it is not
affected by residual B-modes 
and gives a non-parametric representation of the matter density at a given point.

Following the approach of \citet{1996MNRAS.283..837S}, we build the
(2$\,$D) mass aperture statistics by computing the aperture densitometry at each point of a grid on the field. The mass density $M_{\rm ap}$ at a given point is given by:
\begin{equation}
\label{eq:massap2}
M_{\rm ap} = \frac{\Sigma_i\,\gamma_{Ti}\,w_i\,Q_i}{\Sigma_i\,w_i\,Q_i}
\end{equation}
where the sum extends over all the background galaxies and $Q$ is a
smoothing-weighting function. We use the window function $Q$ that maximises
the total signal-to-noise ratio (S/N)  for a SIS \citep{1996MNRAS.283..837S}:
\begin{equation}
\label{eq:qdef}
\begin{array}{lclr}
Q(x) & = & 6\pi\,x^2\times(1 - x^2) & {\rm ; x<0}\\
 & = & 0 & {\rm ; x\ge 0}
\end{array}
\end{equation}
 where: $x=r/r_{\rm an}$, $r$ being the distance to the center of the
 annulus, and \mbox{$r_{\rm an}=5\,$arcmin} (equivalent to \mbox{$=1500\,$pixels}
 \mbox{$=1\,$Mpc}) is the external radius of the annulus. The only
 free parameter, ie. the filter scale $r_{\rm an}$, is chosen as a compromise
 between different needs. On one side, $r_{\rm an}$ needs to be as small as
 possible to avoid to wash out small scale structures. On the other
 hand the annulus needs to be large enough to encompass a significant
 number of background galaxies (typically $1000$) everywhere on the field, in order to get a good and spatially stable S/N. 
Figure \ref{fig:galdist} shows the mass aperture density map of the
 \mbox{$\sim 100\,$arcmin$^2$} patch centered on \adeu , together with the cluster center of mass
 defined by equation \ref{eq:masscenterdef}. The mass
 aperture statistics show the elongated structure of the mass
 distribution. The mass distribution is far from the circular symmetry
 as assumed by SIS and NFW profile fits. This explains why these fits give only a rough estimate of the total mass, as discussed in section \ref{sec:conclusions}.


\section{Comparisons with other observations}


\subsection{Comparison with the galaxy distribution}
\label{sec:compargaldistrib}
\begin{figure*}
\begin{center}
\includegraphics[scale=0.85]{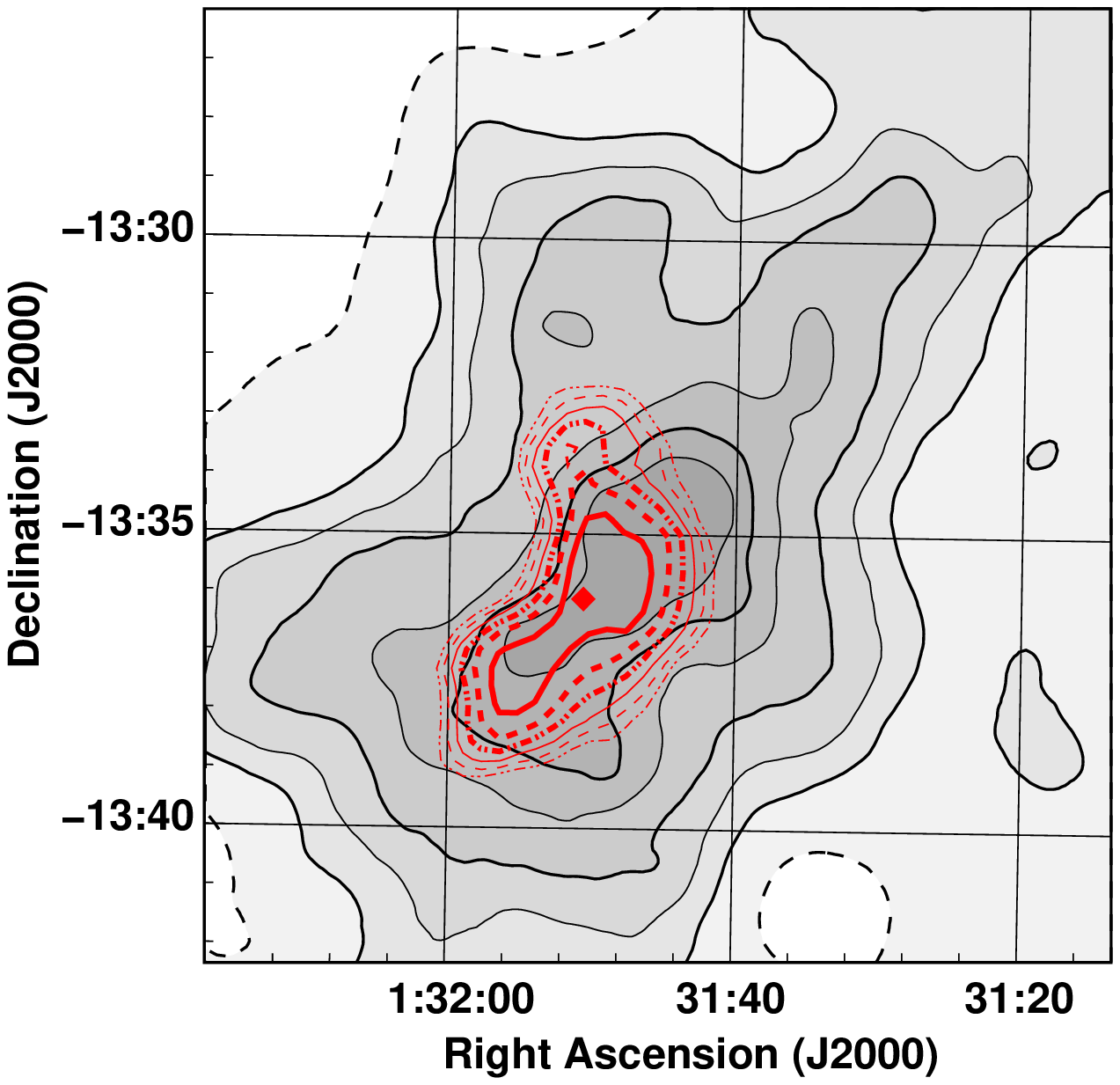}
\caption{Comparison of the weak lensing mass reconstruction with the
  galaxy distribution for \adeu . The black contours represent the isodensity
  contours from the M$_{ap}$ statistics, corresponding to mass
  densities $5\times10^{13}{\rm M}_{\sun}$\,arcmin$^{-2}$
  (thin dot-dashed curve) to $1\times10^{14}{\rm M}_{\sun}$\,arcmin$^{-2}$
  (tick solid curve), with each contour separated by
  $1\times10^{13}{\rm M}_{\sun}$\,arcmin$^{-2}$. The cluster center of mass as
  defined by Equation 14 is indicated by the red diamond.
The grayscale filled
  black contours represent the surface density of $R<23.0$ galaxies
  (background-corrected), corresponding respectively to 0 (dashed
  curve) 2 (solid), 3.3, 5,
  7.5, 10.0 and 12.5 cluster \mbox{galaxies.arcmin$^{-2}$}.}
\label{fig:galdist}
\end{center}
\end{figure*}

Figure~\ref{fig:galdist} shows the galaxy distribution (shown as the
grayscale-filled black contours) as
represented by the surface density of $R<23.0$ cluster galaxies
\citep[ie. after correcting for background/foreground galaxies, see][]{2004A&A...425..783H}. The galaxy
distribution is strongly elongated in the same SE-NW axis as observed
for the weak-lensing reconstructed projected mass distribution. The center of
the galaxy distribution appears somewhat offset to the NW by about
$1^\prime$ with respect
to the center of mass obtained from the weak lensing reconstruction. 
There is also a substructure about $5^\prime$ to the North of the
cluster center, which could be connected with the observed
substructure in the reconstructed mass distribution that extends
northwards from the central mass concentration. 

The internal dynamics of the cluster was studied by \citet{2003A&A...397..431M} through a spectroscopic survey of 112 cluster members. A
high value of the line-of-sight velocity dispersion was found, with
\mbox{$\sigma_{\nu}=1250_{+84}^{-98}$\,km\,s$^{-1}$} after removing
seven interlopers. Assuming dynamical equilibrium, this value of $\sigma_\nu$ leads to a
virial radius of \mbox{$R_{vir}\sim 3.28\pm0.55\,h_{70}^{-1}\,$Mpc} and a virial mass of \mbox{$M(<R_{vir})=3.02^{+0.86}_{-0.89},\times\,10^{15}\,h_{70}^{-1}\,{\rm M}_\odot$} in a $\Lambda$CDM model,
with $\Omega_{m}$=0.27 and $\Omega_{\Lambda}$=0.73. We report this value
in Figure \ref{fig:fittedmass}, for a direct comparison with the results of the NFW and SIS analyses showed in this paper. The
difference in the estimates of the virial mass and the virial radius
obtained by the kinematics and weak lensing could be due to the presence
of substructures which results in an over estimate of the velocity
dispersion. Another source of uncertainties in the kinematics could be the
anisotropy parameter of the cluster velocity distribution. On the other
hand, the lensing values could be biased because of the elongation of the
cluster mass distribution (see section \ref{sec:conclusions}).

\begin{figure*}
\resizebox{\hsize}{!}{\includegraphics{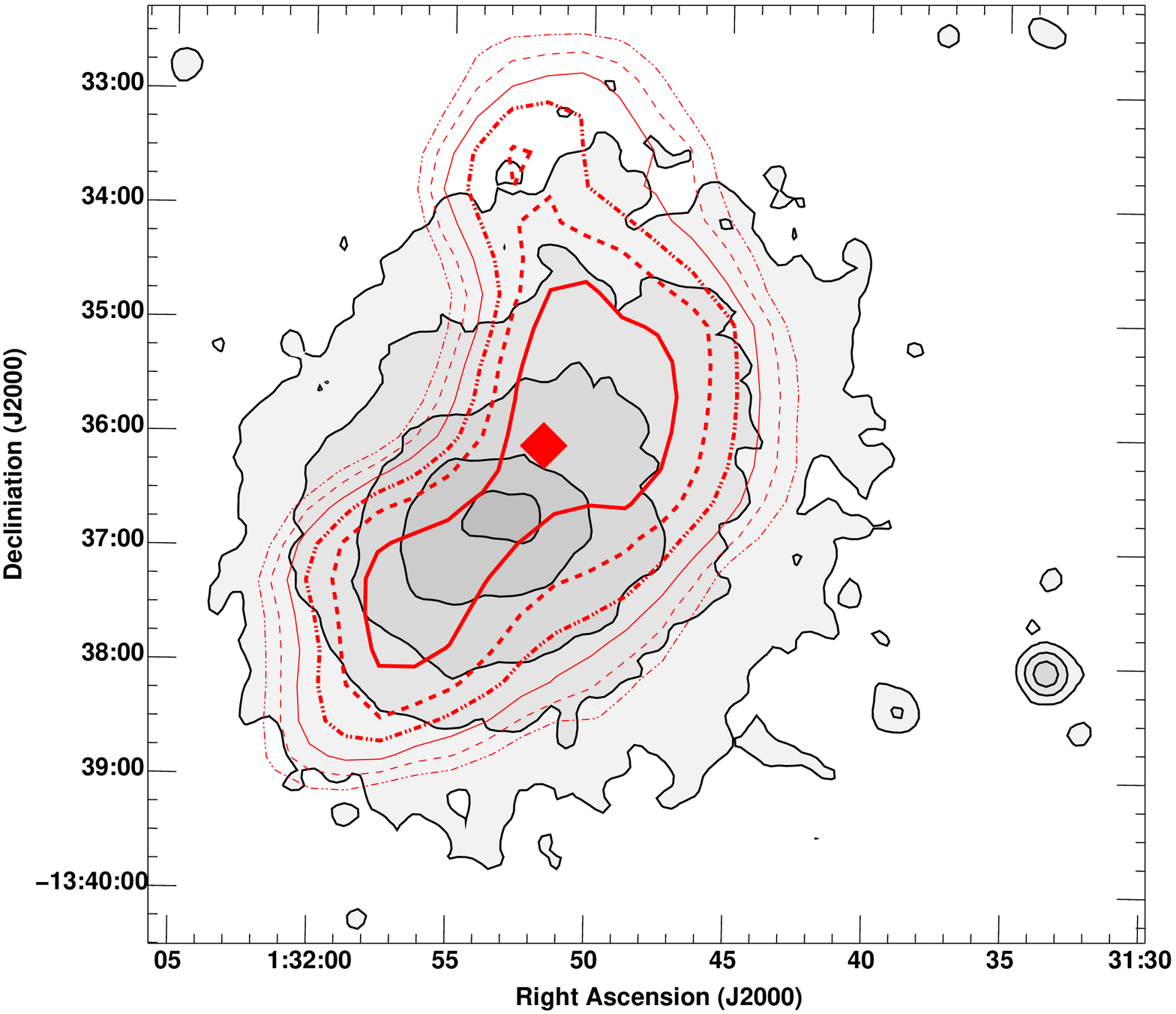}}
\caption{Comparison of the weak lensing mass reconstruction with the
  X-ray emission for \adeu . The red contours represent the isodensity
  contours from the M$_{ap}$ statistics, as in Fig.~\ref{fig:galdist}. The black
  grayscale-filled contours represent the X-ray emission based on the
  XMM imaging. The contours are logarithmically spaced, adjacent
  contours indicating a factor two change in flux density.}
\label{fig:xmm}
\end{figure*}
Evidence in favour of the cluster undergoing a dynamical evolution is
found in the form of a velocity gradient acting along a SE-NW axis,
which is the same preferential direction found from the elongation in
the spatial distribution of galaxies, as well as that
of the cD galaxy. There is also significant deviation of the velocity
distribution from a Gaussian, with evidence for two secondary clumps
at \mbox{$z=0.199$} and \mbox{$z=0.215$}, which appear spatially
segregated from the main cluster. These all indicate that \adeu{} is
undergoing strong dynamic evolution with the merging of two or more
sub-clumps along the SE-NW direction.
\subsection{Comparison to X-ray emission}

The X-ray data are taken from the XMM science archive (Prop \#8423,
PI. J.-P. Kneib, see \citet{2003SPIE.4851..208M} for an analysis). The observations were made in Jan 2001, with an
exposure time of 20\,ksec. The EPIC MOS1, MOS2 and pn images were
combined over the temperature range \mbox{$0.5 - 12\,$keV} and the resulting spatial
distribution of the X-ray emission is shown in Fig.~\ref{fig:xmm} by the
grayscale-filled contours. The X-ray emission is centered on the cD
galaxy (\mbox{$\alpha=$1h31m52.5s}, \mbox{$\delta=-13^\circ 36^\prime 40^{\prime\prime}$}, \mbox{$z=0.2097$}), making
it slightly offset \mbox{($36^{\prime\prime}$)} from the center of mass determined from the weak
lensing by equation \ref{eq:masscenterdef}. The X-ray emission is
elongated along the same SE-NW direction as seen for the weak lensing reconstructed
mass distribution, the emission being most extended
towards the NW. There is no evidence of excess X-ray emission from the
substructure seen in the weak-lensing reconstruction \mbox{$\sim3^\prime$}
to the North of the cluster center.\\
\noindent
From an analysis of a 10\,ksec Chandra ACIS-I (0.3--10\,keV) X-ray observation of the
cluster, \citet{2003A&A...397..431M} obtained a best-fitting temperature of
\mbox{T$_{X}=10.2_{-1.2}^{+1.4}$\,keV} which, assuming $\beta_{spec}=1$, would
correspond to \mbox{$\sigma_v\sim1300\,$km.s$^{-1}$}, and is consistent with the high value of
$L_X(0.1-2.4\,$keV)$ \sim 14\;h_{50}^{-2}\;10^{44}\,$erg.$s^{-1}$ \citep{1996MNRAS.281..799E}. This value of the
velocity dispersion produces a virial mass estimate \mbox{of $3.3\times 10^{15}\,{\rm M}_\odot$}\\
\noindent The mass estimates obtained through our weak lensing analysis are lower than those based on the X-ray temperature and
galaxy velocity dispersions. In a weak lensing analysis of 35 X-ray
luminous clusters at \mbox{$0.15<z<0.30$}, \citet{2006ApJ...653..954D} finds a large
scatter in the relation between the weak lensing mass estimates and
the X-ray luminosity, producing a mass uncertainty of \mbox{$\sigma_{M}=0.44$\,dex}. In a weak lensing study of 24 X-ray luminous
clusters at \mbox{$0.05<z<0.31$}, \citet{2004ApJ...613...95C} found that on average the mass estimates
based on X-ray temperatures and velocity dispersions were 13--27\%
higher than those from the weak lensing analysis. In particular they
found the discrepancy to be much greater for the most massive clusters
(\mbox{$T_{X}>8$\,keV} or \mbox{$\sigma_{\nu}>1122$\,km.s$^{-1}$}), where the mass
excess from the X-ray temperatues or velocity dispersions were
40--75\%. They found that the discrepancy was largest for the two
clusters with the largest X-ray temperatures \mbox{($T_{X}\sim13$\,keV)} and
velocity dispersions, which were known to be undergoing a merging
event, and are far from equilibrium. The high X-ray temperatures would
then be probably due to recent shocks, and the high velocity
dispersions due to substructures and the complex dynamical situation.

\begin{figure*}
\resizebox{\hsize}{!}{\includegraphics{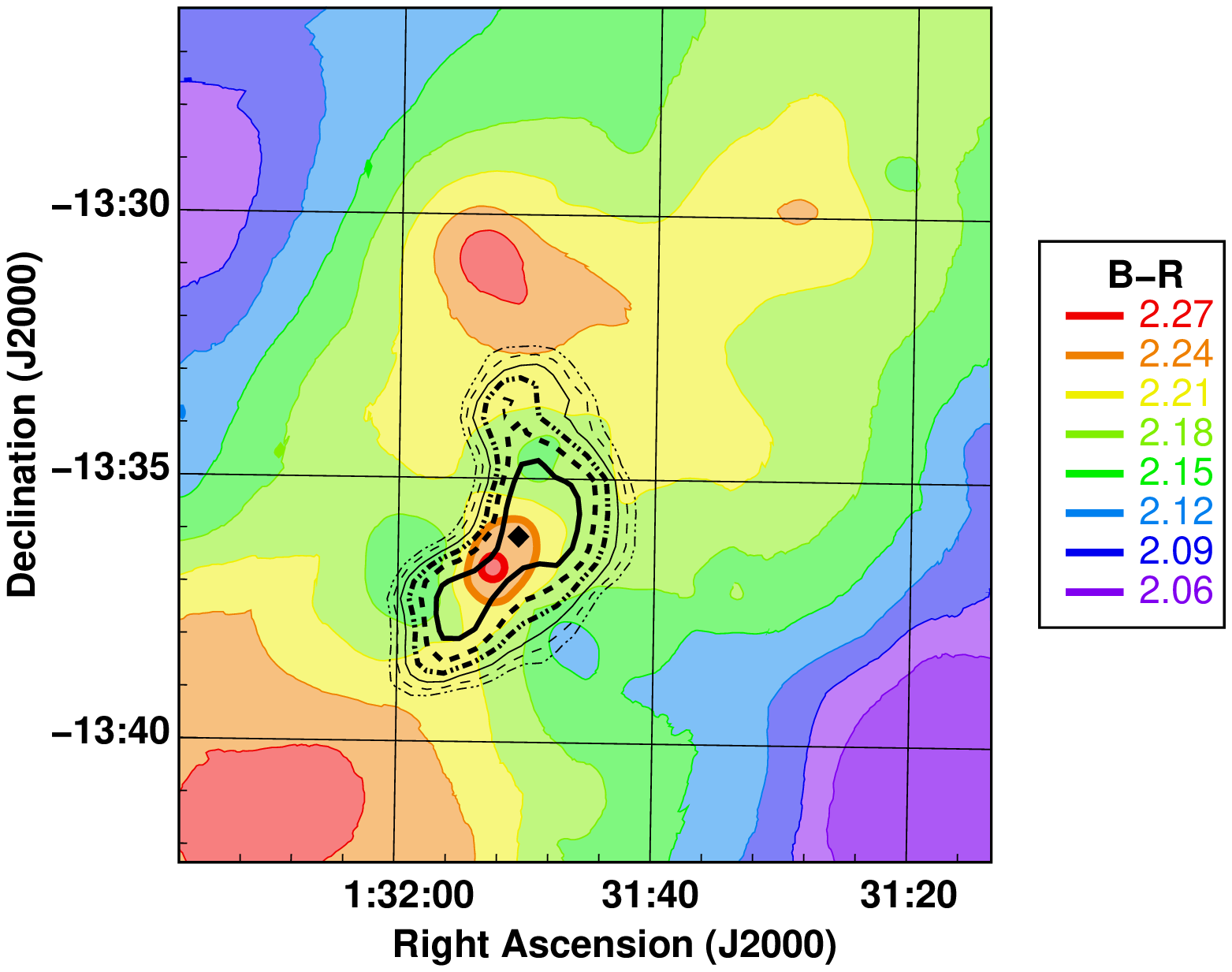}}
\caption{Comparison of the weak lensing mass reconstruction with the
  mean colours of galaxies across \adeu . The red contours represent the isodensity
  contours from the M$_{ap}$ statistics, as in Fig.~\ref{fig:galdist}. 
The coloured contours indicate the mean $B-R$ colour of the $R<21$ cluster
  galaxy population (after statistically correcting for field
  contamination) as a function of spatial position.}
\label{colours}
\end{figure*}

\subsection{Comparison with Galaxy Colours}

The star-formation history of galaxies is known to correlate strongly
with their local environment. In Figure~\ref{colours} we compare the
mass distribution with the mean \mbox{$B-R$} colour of the \mbox{$R<21$} cluster
galaxy population as a function of spatial position. Each galaxy is
weighted according to the probability that it belongs to the cluster,
and then the mean colour of cluster galaxies calculated as a function
of spatial position using an adaptive kernel method. This analysis is
described in details in \citet{2004A&A...425..783H}: the mean galaxy colours (and hence their star-formation
histories) are strongly correlated with the dynamical state of the
cluster, with an alignment of the colours with the main SE-NW
axis. The region with the reddest mean galaxy colours, and hence
the oldest stellar populations, is found at the cluster center of
mass. There are also two regions of red galaxies outside the cluster core that are aligned with the dark
matter distribution, confirming that galaxy evolution is strongly
dependent on the hierarchical build up of clusters through mergers.
Given the uncertain effects cluster dynamics have on the X-ray
emission and galaxy velocity distributions, maps of the mass
distribution based on weak lensing analyses provide an important tool
for understanding the relation between galaxy evolution and the
underlying dark matter distribution \citep{2004MNRAS.347L..73G}.


\section{Conclusions}
\label{sec:conclusions}
We have performed a weak lensing analysis of the galaxy cluster
\adeu{} through a new implementation of the KSB+ algorithm (the OACt
pipeline), and we have also performed a mass reconstruction using Mass
Aperture and parametric statistics. We clearly find a measurable weak lensing signal, and the comparison with optical and X-ray data for this
cluster brings some interesting conlusions.
\noindent First, the centers of the X-ray emission, dark matter, and galaxy
distribution all appear offset from one another, with the center of
mass found from the weak lensing analysis lying between that of the
X-ray and galaxy
distributions, and all the three centers of mass aligned on the main SE-NW
axis of the cluster. Such an effect is seen for the more extreme
Bullet cluster \citep{2004ApJ...604..596C}, and appears to reflect the
different responses of the gas and dark matter components to the merger,
briging a further hint at a cluster merging scenario for \adeu .
\noindent Second, we confirm that \adeu{} is a massive cluster, 
although the mass estimated by weak lensing is lower than the
estimates obtained by \citet{2003A&A...397..431M} from the analysis of
the dynamical properties of the galactic population (assuming the
dynamical equilibrium). On the weak lensing side, there are two
sources of error not taken into account in this
analysis, that could explain this discrepancy. First, the 2D-mass
distribution of the cluster is not circular and this fit of a
circular profile is possibly not accurate. Second, we have not taken
into  account the uncertainty on the critical surface density
(equations \ref{eq:sigmacritdef} and \ref{eq:sigmacritval}), which has
been computed according to the single-source plane approximation at
$z=1$. Considering these uncertainties, the agreement we find among the
different mass estimates should be regarded as satisfactory.


\section*{Acknowledgments}
The work of S. Paulin-Henriksson, V. Antonuccio-Delogu and U. Becciani
has been partially supported through the EC {\em Transfer of
  Knowledge} Marie Curie grant
MTKD-CT-002995, project: {\em COSMOCT}, EC-VI Framework Programme for
R\&D. 
C. P. Haines acknowledges the financial supports provided
through the European Community's Human Potential Program, under
contract HPRN-CT-2002-0031 SISCO. This work is partially supported by
the Italian Ministry of Education, University and Research (MIUR)
grant COFIN200420323 and by the INAF grant ``PRIN 2005''. 
S. Paulin and V. Antonuccio-Delogu would like to thank N. Kaiser
for the online version of IMCAT, A. R\'efr\'egier and D. Clowe for useful suggestions.

\bibliographystyle{aa}
\bibliography{ABCG209-paulin-et-al-astroph-v2}

\end{document}